\documentclass[sigconf]{acmart}
\usepackage{CJK}
\usepackage{xcolor}
\usepackage{balance}

\AtBeginDocument{%
  \providecommand\BibTeX{{%
    \normalfont B\kern-0.5em{\scshape i\kern-0.25em b}\kern-0.8em\TeX}}}

\newcommand\yuling[1]{\textcolor{black}{#1}}
\newcommand\review[1]{\textcolor{black}{#1}}

\newcommand\finalmark[1]{\textcolor{black}{#1}}
\newcommand\ying[1]{\textcolor{black}{#1}}

\copyrightyear{2024}
\acmYear{2024}
\setcopyright{acmlicensed}\acmConference[CHI '24]{Proceedings of the CHI Conference on Human Factors in Computing Systems}{May 11--16, 2024}{Honolulu, HI, USA}
\acmBooktitle{Proceedings of the CHI Conference on Human Factors in Computing Systems (CHI '24), May 11--16, 2024, Honolulu, HI, USA}
\acmDOI{10.1145/XXXXXXX.XXXXXXX}
\acmISBN{XXX-X-XXXX-XXXX-X/XX/XX}

\begin{document}
\begin{CJK*}{UTF8}{gbsn}

\title[Unpacking ICT-supported Social Connections and Support of Late-life Migration]{Unpacking ICT-supported Social Connections and Support of Late-life Migration: From the Lens of Social Convoys}

\author{Ying Lei}
\affiliation{%
  \institution{East China Normal University}
  \city{Shanghai}
  \country{China}
  }
\email{10195102413@stu.ecnu.edu.cn}

\author{Shuai Ma}
\affiliation{%
  \institution{Hong Kong University of Science and Technology}
  \city{Hong Kong}
  \country{China}
  }\email{shuai.ma@connect.ust.hk}

\author{Yuling Sun}
\authornote{Corresponding author.}
\affiliation{%
  \institution{East China Normal University}
  \city{Shanghai}
  \country{China}
  }
\email{ylsun@cs.ecnu.edu.cn}

\renewcommand{\shortauthors}{Ying Lei, et al.}

\begin{abstract}

Migration and aging-related dilemmas have limited the opportunities for late-life migrants to rebuild social connections and access support. \finalmark{While research on migrants has drawn increasing attention in HCI, limited attention has been paid to the increasing number of late-life migrants. 
This paper reports a qualitative study examining the social connections and support of late-life migrants. In particular, drawing on the social convoy model, we pay specific attention to the dynamic changes of late-life migrants' social convoy, the supporting roles each convoy plays, the functions ICT plays in the process, as well as the encountered challenges and expectations of late-life migrants regarding ICT-supported social convoys. 
Based on these findings, we deeply discuss the role of the social convoy in supporting more targeted social support for late-life migrants, as well as broader migrant communities. Finally, we offer late-life migrant-oriented design considerations. }

\end{abstract}

\ccsdesc[500]{Human-centered computing~Empirical studies in HCI}

\keywords{Late-life Migrant, ICT, Social Convoy Model, Social Connection, Social Support, Qualitative Research, China, Older Drifter}

\maketitle

\section{Introduction}


\textbf{\textit{Late-life migrant}} refers to the population who relocates to a different place in the later stages of life, with various reasons of, for instance, family reunification and increased refugee admissions \cite{montayre2017moving, maleku2022we}, looking for better life conditions \cite{du2023understanding, montayre2017moving}, providing care assistance for grandchildren \cite{da2015later, zhou2012space}, dramatic social and political changes \cite{heikkinen2013transnational}, etc. It includes within-national migration \cite{UNFPA2019report, Xinhua2018report} and transnational immigration \cite{maleku2022we, du2023understanding}. With the accelerated processes of urbanization and globalization in contemporary society, late-life migration has become a burgeoning phenomenon worldwide. Migration has a significant impact on people's quality of life and social relationships \cite{foroughi2001relationships, ryan2011migrants}. Particularly, compared to those who migrate as younger people, late-life migrants often confront challenges of both migration and aging in their new places of resettlement \cite{litwin2008late}. Their lives are often isolated \cite{liu2016dilemma}, with poor mental health \cite{tang2022social, wang2022mental} and low-level life quality and well-being \cite{da2015later, liu2019nothing}.


Social connection has been shown to be an important part of health and well-being for old adults \cite{liu2020can, van2017subjective}. However, migration threatens late-life migrants' social connections built over the years and the accessibility to social support \cite{heikkinen2013transnational}. The risk of losing social connections is significant for late-life migrants due to their conditions of retirement and decline in physical functions \cite{hawkley2008social}. Meanwhile, migration in late life means the lack of work or education opportunities in the new environment \cite{heikkinen2013transnational}, and more difficulties in relocating to the new culture \cite{da2015later}, which further hinders them from rebuilding new social connections. Given these barriers, understanding the social connections and challenges faced by late-life migrants, and subsequently formulating targeted strategies to assist them in rebuilding social connections and improving the quality of their migrant life, has become an extremely important issue.

Information and Communication Technology (ICT) has played an extremely important role in the process of restructuring migrants' social connections, assisting them in maintaining contact with old relationships and connecting with new ones \cite{hiller2004new, WangYan2019MobileConnectivityand}. In Human-Computer Interaction (HCI) and Computer Supported Cooperative Work (CSCW), researchers have paid increasing attention to research on ICT-supported social relationship reconstruction among various kinds of migrants (e.g., refugees\cite{xu2017community, hussain2020infrastructuring}, immigrants \cite{rao2016asian, hirsch2004speakeasy, wyche2012we}, migrant workers and their children \cite{pan2013exploration, liu2014enriching, gan2020connecting}). To our knowledge, research has yet to focus on how ICT helps or impacts the social connection and challenges faced by late-life migrants. 
\finalmark{This research gap is critical because researchers in both academic and governmental communities need an in-depth understanding of how to appropriately implement ICT-based social support systems to assist late-life migrants in overcoming the challenges they encounter in a new environment and enhance their quality of social life. }

Our study addresses these research questions through a qualitative study of 13 late-life migrants in China. In particular, drawing on Kahn and Antonucci's \textit{social convoy model} \cite{rl1980conboys}, we unpack the social connections and support of late-life migrants through the lens of social convoy.
\finalmark{Our findings elaborate on the structure and transitions of late-life migrants' social convoys, the roles of ICT in this process, as well as the perceived challenges and expectations of late-life migrants. Based on the findings, we deeply discuss the role of the social convoy in supporting more targeted social support for late-life migrants, as well as broader migrant communities.}

\finalmark{
Our paper makes the following contributions to HCI: (1) we contribute to previous studies examining the social support of migrants by offering a qualitative, empirical study that examines the social connections and support of late-life migrants, which is largely ignored in existing literature. 
(2) We use the model of social convoy to frame and analyze our empirical findings, discussing the broader implications of enabling social convoy in migrant communities. This contributes a new perspective for understanding the social connections and support of migrants on the one hand, and informing more targeted support strategy designs on the other hand. 
(3) Drawing on our findings, we offer late-life migrant-oriented design considerations for better social support for them. 
}

\section{Background and Related Work}

In this section, after the study background of late-life migrants in the world and China, and the theoretical background of the social convoy model, we will review multiple strands of literature to situate our work, including the social connections of migrants, older adults, and late-life migrants.

\subsection{\finalmark{Study Background: Late-life Migrants in the World and in China}}

\finalmark{
Population mobility is a very common social phenomenon. People move from the place of their birth to a new city for education, employment, and to better life \cite{liu2014enriching, liu2017subjective}. In this trend of population mobility, older adults are traditionally considered a less mobile population due to age-related challenges. However, with the urbanization and development of contemporary society, a growing trend of late-life migration globally has been identified due to various reasons, including seeking better healthcare \cite{du2023understanding, montayre2017moving}, joining family members \cite{da2015later, zhou2012space}, or pursuing retirement in a different environment \cite{warnes2006older}.}

\finalmark{
In China, influenced by the traditional family culture \cite{sun2010value}, increasingly older adults have begun to migrate along with their children for helping alleviate children's life burdens, familial reunions, or requiring care from their children, who cannot relocate to where they reside~\cite{liu2019nothing, tang2022social}. As a kind of late-life migrant, these people are called \textit{`older drifters'} (老漂, \textit{lao piao}) in China. 
With the acceleration of social mobility among young people, the phenomenon has become a prominent social phenomenon in China. Newly released census data show that the population of late-life migrants in China has increased from 5 million to 18 million in the past two decades, among whom 77\% are older drifters \cite{Xinhua2018report, UNFPA2019report}. 
The older drifters have drawn significant attention from social scholars \cite{tang2022social, liu2023association, bao2022embedding}. They discussed social adaptations, community integration, mental health, and life quality of older drifters, and called for efforts from social workers, communities, and governments to better support late-life migrants. 
Our work is situated in this particular context of the older drifter phenomenon in China. }

\subsection{Theoretical Background: Social Convoy Model}

\begin{figure}[ht]
  \centering
  \includegraphics[width=0.85\linewidth]{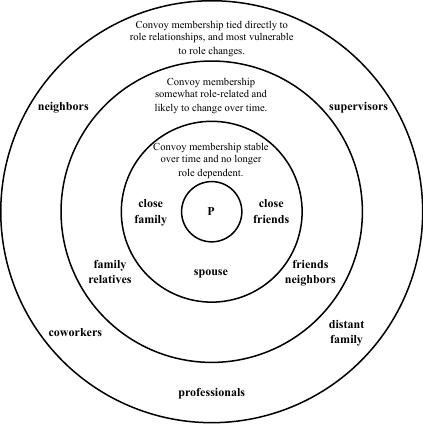}
  \caption{\yuling{Hypothetical example of a convoy \cite{rl1980conboys}.}}
  \label{Original Social Convoy Model}
  \Description{It represents a typical example of social convoy, in which three concentric circles are used to represent the social convoy of one person (P), and members in these circles are people who provide support to P. The innermost circle consists of people who are closest to P, while the outermost circle represents members who are least close to P. The bold letters present the members in each cycle, while the smaller letters in the upper position of each cycle explain the features of members in this cycle. }
\end{figure}

\review{Kahn and Antonucci developed the convoy model of social connections \cite{rl1980conboys}, highlighting social connections' protective function as individuals experienced age-related events throughout their lifetime \cite{nestmann2012social, schaie2010handbook, baltes2014life}. According to this model, individuals are surrounded by supportive others who move with them throughout the life course. Figure \ref{Original Social Convoy Model} represents a hypothetical example of the social convoy, in which three concentric circles are used to represent the social convoy of one person (P), and members in these circles are people who provide support to P. The innermost circle consists of people who are closest to P, while the outermost circle represents members who are least close to P. 
These relationships vary in their closeness, quality, function, and structure, which are influenced by personal and situational characteristics \cite{ajrouch2001social, ajrouch2005social} while having significant implications for health and well-being. According to \cite{rl1980conboys}, social convoys move with individuals dynamically across time and the life course.}

\review{As a theoretical basis, the social convoy model has been widely used to examine the social support of people in different socioeconomic and cultural contexts \cite{antonucci2014convoy} and life stages \cite{nestmann2012social, schaie2010handbook, baltes2014life}. 
Particularly, researchers have recognized the importance of social convoy in improving older adults' quality of social life \cite{fuller2010theoretical, fuller2020convoy}. For instance, \citet{fuller2020convoy} utilizes the social convoy model to examine family relationships in later life, highlighting the advantages and inclusiveness of this model in guiding present and future research to tackle the challenges older adults encounter. Studies have also demonstrated that the best way to protect older adults against social loneliness is to develop a social convoy that accompanies them across the life span \cite{rl1980conboys}.}

\finalmark{
We consider the social convoy model as an appropriate theory for understanding late-life migrants' dynamic social support. For migrants, when they migrate to new places, their convoys in the old places would be disrupted, leading to the `broken convoy' effect \cite{park2015empirical}. 
Understanding late-life migrants' social support through the lens of social convoys will help theoretically frame the social support conditions of late-life migrants on the one hand, and 
present the dynamics of the support during the migration process on the other hand, which can provide critical insights for long-term and sustainable support for late-life migrants \cite{fuller2010theoretical}.
Moreover, while recent studies have started using the social convoy model to understand late-life migrants \cite{park2015empirical,maleku2022we}, most primarily analyze the influence of different factors, without considering the specific roles of ICT on this particular group. Our study adds to this literature by particularly examining the role of ICT in mediating late-life migrants' social convoy, which enriches the applications of the social convoy model. }

\subsection{Social Connections of Migrants} 

\review{
Research on migrants, particularly immigrants and refugees, has drawn increasing attention in recent HCI. Existing work has demonstrated that maintaining old and creating new social connections are significant but challenging for migrants \cite{heikkinen2013transnational}. 
Given the challenges facing migrants (e.g., resettlement and social capital \cite{almohamed2019rebuilding, hsiao2018technology}, community building \cite{rao2016asian}, health \cite{brown2014reflection, tachtler2020supporting}, child care and education \cite{wong2019parenting, brown2012takes}, culture adoption \cite{sabie2020memory}, communication \cite{liu2014enriching, gan2020connecting}, etc., researchers have proposed a lot of technological solutions, such as ICT-supported remote family communication \cite{pan2013exploration, gan2020connecting}, local community building and engagement \cite{xu2017community, hussain2020infrastructuring}, etc. For instance, \citet{gan2020connecting} focuses on mobile video calling between `left-behind' children and their migrant parents, and reveals a key practice of `facilitation work', performed by grandparents, in this process. In a different context, \citet{rao2016asian} describes how a specific Facebook group, the Asian American Chicago Network (AACN), helps immigrants in the U.S. utilize social media to build a local community.} 

\review{
Recently, with the constant promotion of globalization, migration is no longer a special case for immigrant and refugee communities but a popular phenomenon among hundreds of millions of people across diverse social, economic, and cultural backgrounds and roles (e.g., international students, migrant workers \cite{liu2014enriching, gan2020connecting}, older migrants \cite{tang2022community, liaqat2021intersectional}). Among them, older adults are traditionally considered as less mobile populations due to age-related issues and thus are less covered in existing research \cite{WangYan2019MobileConnectivityand}. With exceptions of \citet{tang2022community} and \citet{liaqat2021intersectional} which explore older migrants and grandparents in multi-generational immigrant families, issues related to late-life migrant and their social support are largely underexplored. }

\review{
To better address the challenges confronted by late-life migrants, a holistic and dynamic portrait is needed \cite{schaie2010handbook, baltes2014life}, instead of only maintaining old social connections \cite{wyche2012we, gan2020connecting} or creating new relationships \cite{xu2017community, hussain2020infrastructuring} from a static perspective. Such a holistic and dynamic portrait can help researchers and developers appropriately implement ICT-based social support systems. Our study adds to this literature by deeply unpacking the social connection and support conditions of late-life migrants in China. }

\subsection{Social Connections of Older Adults}

\review{Social connections and support are crucial for older adults. Helping maintain and rebuild social connections of older adults, which has been proven to be effective in reducing older adults' social isolation \cite{liu2020can} and promoting their quality of life and well-being \cite{van2017subjective, mitchell2004aging}, is always a research focus in HCI and related fields. 
Researchers have developed various digital tools (e.g., \cite{brewer2017xpress, baker2019interrogating}) to promote social participation and engagement of older adults, and simulate their social connection, through maintaining their existing social relationships with, e.g., relatives \cite{cornejo2013ambient}, children\cite{gutierrez2016mom}, and caregivers \cite{zubatiy2021empowering}, and building new relationships \cite{brewer2017xpress}.}

\review{
However, most existing studies are conducted with older adults who live in their familiar environments, which are very different from the late-life migrants this study examines. 
Specifically, late-life migrants move to a new social environment wherein their social connections are highly complex, encompassing both their original social relationships and new ones \cite{heikkinen2013transnational}. Compared to older adults aging in place or other kinds of older migrants who spent their adult life in the host places and are older now \cite{maleku2022we, du2023understanding}, late-life migrants move to the new place at an older age, which means more difficulties in maintaining their old relationships and rebuilding new social support network \cite{baldassar2017aging} in the host communities. Yet, despite the increasing numbers of late-life migrants, very little is published in HCI on this population \cite{maleku2022we, walters2002later}. }

\review{In particular, previous studies have shown that later-life migrants often suffer significant challenges in maintaining old connections and creating new connections due to the challenges in culture adaptation, less work and education opportunities \cite{heikkinen2013transnational, hawkley2008social, da2015later}, which further causes a series of negative effects, e.g., social isolation \cite{liu2016dilemma}, poor mental health \cite{tang2022social, wang2022mental}, low-level life quality and well-being \cite{liu2023association, da2015later}, etc. Moreover, compared to younger migrants, late-life migrants need to deal with the aging-related losses and limitations they face (e.g., digital divide, physical decline \cite{friemel2016digital, barbosa2019can}), which further exacerbate these challenges. Our study provides additional insight into the social connections and support of older adults in HCI by examining late-life migrants. Further, HCI scholars have suggested ICT for older adults should go beyond standard accessibility or `senior-friendly' considerations, adopting more holistic approaches to design for older adults, such as their generational context \cite{tang2022never, sun2014being}, social ecology \cite{sun2014being, sun2015reliving} and value systems \cite{sun2015reliving}. Echoing this effort, our study examines ICT-supported social support system design by enabling the dynamic social context of late-life migrants.
}

\section{Methodology}

\review{
Our study aims to examine late-life migrants' social connections and support during and after migration, ICT's roles in this process, and their encountered challenges and expectations to further social supporting systems. To this end, we conducted semi-structured interview studies with 13 late-life migrants in China. We now will elaborate on the methodology of our study. }

\begin{table*}[htbp]
  \caption{Demographic information of participants.}
  \label{tab:demographic information}
  \begin{tabular}{llllllll}
    \toprule
    \yuling{ID} & \yuling{Gender} & \yuling{Age} &	\yuling{Location} & \yuling{Hometown} & \yuling{Duration} & \yuling{Motivation} & \yuling{Key ICT Tools}\\
    \midrule
        \yuling{P1} & \yuling{Female} & \yuling{71-75} & \yuling{Shanghai} & \yuling{Shandong} & \yuling{20 years} & \yuling{Grandchildren} & \yuling{Tel., WeChat}\\
        \yuling{P2} & \yuling{Female} & \yuling{51-55} & \yuling{Shanghai} & \yuling{Shandong} & \yuling{6 months} & \yuling{Grandchildren} & \yuling{Tel., WeChat, Tencent Meeting, TikTok, Headlines}\\
        \yuling{P3} & \yuling{Female} & \yuling{61-65} & \yuling{Shanghai} & \yuling{Henan} & \yuling{4 years} & \yuling{Grandchildren} & \yuling{Tel., WeChat}\\
        \yuling{P4} & \yuling{Male} & \yuling{66-70} & \yuling{Shanghai} & \yuling{Henan} & \yuling{9 months} & \yuling{Grandchildren} & \yuling{WeChat, Headlines}\\
        \yuling{P5} & \yuling{Female} & \yuling{66-70} & \yuling{Shanghai} & \yuling{Anhui} & \yuling{10 years} & \yuling{Grandchildren} & \yuling{WeChat, TikTok, Tencent Meeting}\\
        \yuling{P6} & \yuling{Female} & \yuling{56-60} & \yuling{Shanghai} & \yuling{Hunan} & \yuling{6 months} & \yuling{Reunion} & \yuling{Tel., WeChat, TikTok}\\
        \yuling{P7} & \yuling{Male} & \yuling{61-65} & \yuling{Shanghai} & \yuling{Shandong} & \yuling{2 years} & \yuling{Grandchildren} & \yuling{Tel., WeChat, TikTok}\\
        \yuling{P8} & \yuling{Female} & \yuling{51-55} & \yuling{Shanghai} & \yuling{Shandong} & \yuling{8 months} & \yuling{Grandchildren} & \yuling{Tel., WeChat, TikTok}\\
        \yuling{P9} & \yuling{Male} & \yuling{56-60} & \yuling{Shanghai} & \yuling{Shandong} & \yuling{1-2 years} & \yuling{Grandchildren} & \yuling{Tel., WeChat, TikTok}\\
        \yuling{P10} & \yuling{Female} & \yuling{61-65} & \yuling{Shanghai} & \yuling{Jiangxi} & \yuling{1 year} & \yuling{Grandchildren} & \yuling{WeChat, TikTok}\\
        \yuling{P11} & \yuling{Female} & \yuling{56-60} & \yuling{Beijing} & \yuling{Shandong} & \yuling{3 years} & \yuling{Grandchildren} & \yuling{Tel., WeChat}\\
        \yuling{P12} & \yuling{Male} & \yuling{66-70} & \yuling{Beijing} & \yuling{Fujian} & \yuling{2 years} & \yuling{Grandchildren} & \yuling{WeChat, TikTok, Headlines}\\
        \yuling{P13} & \yuling{Female} & \yuling{61-65} & \yuling{Beijing} & \yuling{Helongjiang} & \yuling{2 years} & \yuling{Grandchildren} & WeChat, TikTok, Headlines, Little Red Book\\
  \bottomrule
\end{tabular}
\end{table*}

\subsection{Participants}
\review{We looked for participants who 1) had retired and 2) had left the place where they originally lived and worked to a new one after retirement, and 3) had lived in the new place for more than 6 months. Considering the definition of `older adult' was often impacted by various social and cultural factors \cite{sun2014being}, we set `retirement' as the criteria of late-life, instead of the general `age greater than 65 years'. We integrated the introduction of ourselves, our intention of doing research about older drifters, interview requirements, commitments to data privacy and safety, and our contact information into a poster, and shared with our classmates, friends, neighborhoods, and some WeChat groups to recruit participants. We also invited recruited participants to help share the poster with their own social network. Through these hybrid methods, we finally recruited 13 older drifters, with details shown in \autoref{tab:demographic information}.}

\review{
Among them, 9 were females and 4 were males, with ages ranging from 51 to 75 and migrating from 8 months to 20 years. They came from seven provinces of China and currently lived with their children in different communities in Shanghai and Beijing, two biggest cities of China. Except for P6, who was for conducting household chores and family reunions, all others migrated to care for grandchildren. The right column lists the technologies participants have used for social connecting, with the detailed descriptions in \autoref{technologies specific to the chinese context} in Appendix.}

\subsection{Data Collection}

\finalmark{
According to the social convoy model~\cite{rl1980conboys}, individuals are surrounded by supportive ones who move with them throughout the life course. The key principle of this model is to illustrate these supportive members according to the closeness with targeted individuals. 
Drawing on this key principle~\cite{rl1980conboys} as well as the aims of this study in examining late-life migrants' social connections and support during and after migration, informing the better supporting interventions, we designed the interview questions around (1) supportive members of older drifters before and after migration, the closeness of these relationships with older drifters and reasons; (2) support from these relationships; (3) the used ICT for social connecting and roles; and (4) encountered challenges and expectations to future ICT-mediated social supporting. All questions were designed to be general so they could be more inclusive to interviewees with various levels of experiences. When some interesting points or prior experiences were mentioned, we followed up for more details and concrete examples. }

\review{All interviews were conducted in Mandarin by two authors in person \yuling{(N=3)} and remotely via Tencent meeting \yuling{(N=3)}, WeChat Voice calls \yuling{(N=3)}, and phone calls \yuling{(N=4)}. Before the interviews, we informed participants of our intention and background information. Each interview lasted for 40 to 60 minutes. With participants' permission, we recorded all interviews and transcribed them into Chinese verbatim after the interviews. All data collected during the study was used in an anonymized way, i.e., there was no link between the collected data and an individual user. PX was used to denote participants.}

\subsection{Data Analysis}

\finalmark{
We applied thematic analysis \cite{braun2012thematic} to analyze all interview data in a deductive approach \cite{patton1990qualitative}. Two authors, who collected the interview data, participated in this analysis process. We first familiarized ourselves with data with the initial analytic interests in older drifters' social connections and supports. We read the collected data back and forth, marking the ideas independently. We had regular meetings on a weekly basis to discuss ideas and ensure reliability. Through this literature process, each of us generated an individual list of initial codes through the whole data. We compared and combined our initial codes through discussion and general the final code lists, including migration motivation, attitudes, experiences, social connection, dynamics, and closeness, social support and stress, digital tool uses, as well as some emerging themes such as `Sandwich Generation' \cite{zhang2006will} and the culture of returning to roots. Based on this initial code list, we re-focused our analysis at the broader level of themes and consolidated our codes into an overarching theme. Through several rounds of discussion, we identified our final thematic map, including (1) older drifters' social convoy structures and transformation before and after migrating; (2) the supporting roles of members in each convoy; (3) the roles of ICT in the convoy; and (4) the encountered challenges of older drifters and their expectations to ICT-mediated social supporting systems.
In the following sections, we present the full details of these themes.
}

\begin{figure*}[ht]
  \centering
  \includegraphics[width=0.90\linewidth, keepaspectratio]{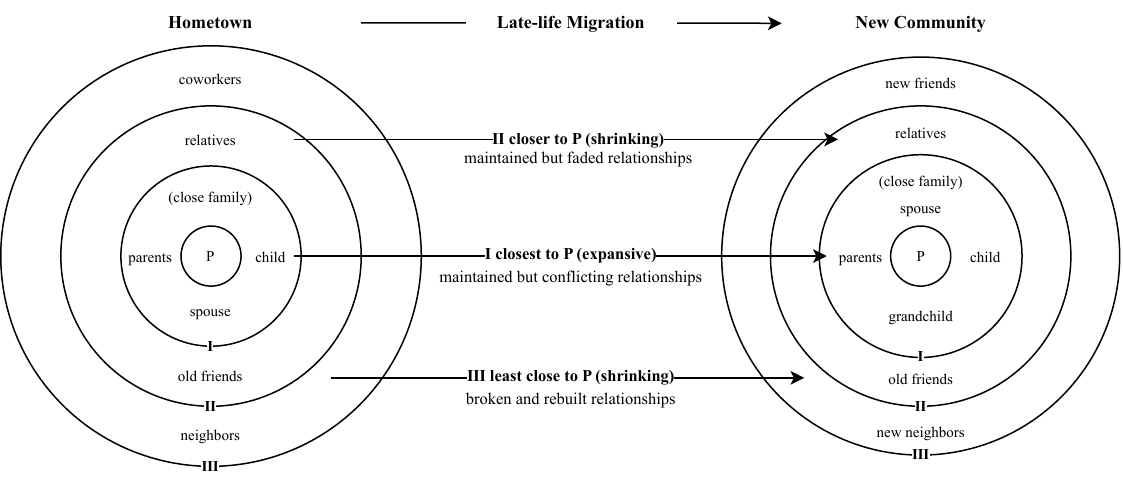}
  \caption{Structure and transitions of older drifters' convoys. Three concentric circles are used to represent close, closer, and the closest relationships respectively (from outer to inner). The left and right parts show the structures of older drifters before and after migration respectively, along with circle scalability and arrows to indicate the transitions. }
  \label{Structure and Transitions of Older Drifters' Convoys}
  \Description{Three concentric circles are used to represent close, closer, and the closest relationships respectively (from outer to inner). The left and right parts show the structures of older drifters before and after migration respectively, along with circle scalability and arrows to indicate the transitions. }
\end{figure*}

\section{FINDINGS}

\review{In this section, we first present the social convoy of older drifters, including structures and transitions before and after migrating and the supporting roles of different convoy members. After that, we elaborate on the roles of ICT in convoys, as well as the encountered challenges and the expectations to ICT-mediated social supporting systems.}

\subsection{Social Convoys of Older Drifters}
\label{structure}

\review{The structure of older drifters' convoys is illustrated in \autoref{Structure and Transitions of Older Drifters' Convoys}. According to the closeness, we divided all relationships into three levels. }

\subsubsection{Closest Relationship}

\review{
The smallest concentric circle in \autoref{Structure and Transitions of Older Drifters' Convoys} consists of people who are very close to P, and remain relatively stable during migrating, despite changes in geographical proximity or communicating frequency. For older drifters in our study, these people were primarily their families, e.g., spouses, parents, children, and grandchildren. 
The migration impacted the memberships in this convoy to some extent. For instance, many of our participants reported that, grandchildren were added into this convoy after migration, while some original relationships were impacted due to the fragmentation between couples (P2-5,7-11,13) and  intergenerational differences (P3,4,6,8).}

\review{For example, many of our participants expressed that the migration caused their separation from their partners. Even some migrated along with their partners, but they often at different times (P3-5,9,13) and with different reasons (P2,6,7,8,10,11). 
The fragmentation couldn't be compensated by the reunion with their children and grandchildren. As P3 said, \textit{`Although my son and daughter-in-law are very good to me, my husband and I were not together at that time. I seemed to have unspeakable grievances'} (P3). Meanwhile, intergenerational differences, such as living habits (P4), location of packing things (P6), and dressing style (P8), also brought potential risks to family relationships. To address this, many of our participants chose to live in different houses with their children. }

\subsubsection{Closer Relationship}
\review{
The middle circle consists of relatively stable relationships that are less dependent on roles that they play in life, such as relatives, co-workers and friends \cite{rl1980conboys}. For older drifters, these members were primarily their relatives and old friends. Our participants expressed that they had accumulated long-term relationships with these members, sharing similar growing experiences, economic conditions, and parentage. }

\review{After migration, these relationships were maintained and were hard to be replaced by new friends in a short time. However, the physical distance and less face-to-face contact caused their limited connections and interactions with these members, which in turn reduced the quality of relationships. 
P1, for instance, had migrated about 20 years. She still maintained relationships with her relatives and old friends in her hometown; however, less in-person contact significantly reduced the intimacy among them. 
}
\begin{quote}
    \review{\textit{`We only selectively chat and help with each other. Many relatives only share the good news with you. For instance, if he suffered from a serious illness, he might only tell me after he recovered.'} (P1)}
\end{quote}

\subsubsection{Least Close Relationship}
\label{challenges2}

\review{The outermost concentric circle represents convoy members who are least close to P, such as supervisors, co-workers and neighbors.  \cite{rl1980conboys}. Their relationships are often not stable and extremely vulnerable to role changes. For our participants, their relationships with these members were almost broken due to migration, replaced by those who shared some common features in the new places, such as identity (e.g., children caregivers (P2,3,11,13), older drifters (P8) and retirees (P2)) and hobby (P4,5,9). Our participants expressed that these common features provided them with a sense of resonance. As P13 told us,}

\begin{quote}
    \review{\textit{`In my community, there are 20 or 30 children caregivers. Many of us used to work, but now we care for our grandchildren and embrace this transformation together. We share a close bond and exchange knowledge and information.'} (P13)}
\end{quote}

\review{Same hobbies, such as fishing (P4), Taichi (P5) and dancing (P6), also facilitated connections and social participation. Our participants hold the view that hobbies helped to bridge cultural barriers (P4) and gain respect and confidence (P5,6), thus promoting connections. As for the means of rebuilding these connections, many of our participants (P2,3,5,8,10,11,13) expressed that they usually went to public spaces in the communities wherein they had more opportunities to meet people with similar features. Additionally, re-employment and re-education were also considered by our participants as good ways to form new connections within this convoy.}

\review{
It was worth noting that, the rebuilding process was full of challenges for older drifters. For instance, language and culture barriers were reported by most participants (P1,2,5,7-9) as one main challenge in rebuilding the connections with local people especially when they moved to a new place at the beginning. Many participants experienced low acceptance by the local people (P1,2,5,8,9,12); Some (P1,5,9,12) even reported their experiences of being discriminated by the locals, such as \textit{`the locals are taking it for granted that bad things are done by outsiders'} (P1), \textit{`non-locals can never belong to the circle of locals'} (P5), `\textit{being ignored or even ridiculed when asking for help'} (P9,12). 
}

\subsection{Support Roles of Older Drifters' Convoy}
\label{role}
\begin{figure}[ht]
  \centering
  \includegraphics[width=\linewidth]{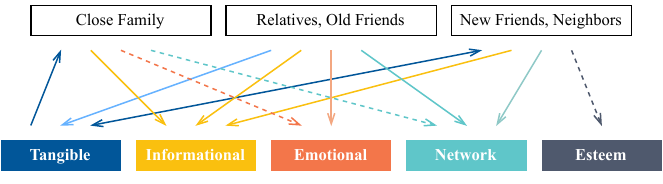}
  \caption{\review{Support roles of different convoy members. Arrows were used to connect older drifters and convoy members: colors indicated corresponding support types, arrow direction meant the relationship of providing/giving/exchange support; lighter colored line represented limited/faded support, and dotted line signified giver-dependent support.}}
  \label{Support Roles of Different Convoy Members}
  \Description{Arrows were used to connect older drifters and convoy members: colors indicated corresponding support types, arrow direction meant the relationship of providing/giving/exchange support; lighter colored line represented limited/faded support, and dotted line signified giver-dependent support.}
\end{figure}

\review{According to the typical social support theory \cite{cutrona1992controllability}, which divided support into five types - tangible, informational, emotional, network, and esteems support, we analyzed the interactions between the members in different convoys and older drifters, and summarized support roles of different convoy members, i.e., which convoy members provide which kinds of support. 
 \autoref{Support Roles of Different Convoy Members} showed the overview of this relationship.}

\subsubsection{Tangible Support}

\review{Tangible Support, known as instrumental support, refers to providing material aid (e.g., money, food, books) or assistance for direct and indirect tasks (e.g., babysitting, transportation, housework, etc.) \cite{coursaris2009analysis}. According to our participants, older drifters mainly provided tangible support to the closest convoy, received limited support from the closer convoy, and exchanged support with the least close convoy. }

\review{Specifically, in the closest convoy, older drifters mainly served as tangible support providers in fulfilling filial piety to their parents and supporting their children physically and financially. Many of our participants stated that this kind of support they `had to' provide(e.g., such as babysitting and housework for their children) was often overwhelming for them and made them breathless, exhausted, and tied down. As most of their time was taken up by these tasks, they may not have time to familiarize themselves with new places (P9), engage in hobbies (P1,3,5), community activities (P3,7), and online connection and activities with the long-distance convoy (P2,8). For example, P1 had been constantly taking care of grandchildren since retirement and she described herself as \textit{`an always running assembly line without any chance to get out or have fun'}.}

\review{In the closer convoy, older drifters mainly receive tangible support from these members, and no experiences were reported by our participants as the support providers. P9, for instance, shared the experience that his old friends in his hometown sent him emergency medicine and essential supplies during the COVID-19 pandemic. Although he had already recovered when the supplies arrived, he was deeply touched by this behavior. This kind of tangible support was relatively rare due to the distance issue.  }

\review{In the least close convoy, older drifters served as both tangible support providers and receivers. For example, P8's son often helped his landlord repair the broken tubes and the landlord gave them vegetables grown by herself in daily life in return; by bringing mats and toys from their own homes, P13 and her neighborhood creatively transformed a public blank place into a `kindergarten' in their community, where they supervised grandchildren, played and chatted together, making babysitting more fun and less troublesome.} 

\subsubsection{Informational and technological Support}

\review{Informational support refers to providing advice/suggestions, guidance, and facts/news to help recipients solve or manage a problem \cite{feng2010influences}. According to our participants, three convoys provided informational support in different information dimensions and content.
For instance, families in the closest convoy and new friends and neighbors in the least close convoy often provided information about the host community and local life, such as public services (e.g., news, shopping, repairs, volunteer contacts) (P9,13), local dialect (P5), local social etiquette (P8), and leisure activities (P2,9), which were highlighted by our participant as useful in avoiding mistakes, promoting adaption to new communities, and helping prevent unnecessary complications. Meanwhile, relatives and old friends in the closer convoy primarily provided information on changes and policies in their hometown. }

\review{In addition, members in the closest convoy (e.g., children, grandchildren, and spouses) also provided necessary technical support for older drifters, helping reduce possible technical barriers in, for instance, online trip booking (P1), online shopping (P3,4,9), video conference operating (P2,5), and online hospital registration (P8,11).}

\subsubsection{Emotional Support}

\review{Emotional support is defined as `specific lines of communicative behavior enacted by one party with the intent of helping another cope effectively with emotional distress' \cite{burleson2003experience}. For older drifters, their emotional support was mainly provided by their close families, relatives, and old friends, while new friends and neighbors were reported to have little influence on emotional support due to their limited duration of contact and emotional accumulation. Many of our participants explained that they actually didn't have very close relationships with new friends and neighbors in the new place, and their interactions were limited to trivial affairs, such as babysitting, fresh things around them, etc. }

\review{Meanwhile, although the closest convoy was often considered the primary emotional support source, it was not always true for all the members in this convoy. Instead, it was provider-dependent. For instance, 
although younger generations could perceive older drifters' troubles and provide some emotional support to some extent (P1,2,6,8,13), many participants reported that they were often emotionally overlooked by their children (P10), and some preferred hiding troubles in front of them to avoid `unnecessary worries' as an additional burden to them (P3-7,9). As P6 shared with us, }

\begin{quote}
    \review{\textit{`My son is very concerned about me, fearing that I might experience negative emotions. But in reality, I hope to take good care of them. I've never expressed my negative emotions to my son, only when it became too much. I worry about the stress young people face and don't want to burden them.' }(P6)} 
\end{quote}

\review{In the absence of emotional support from new relationships and uncertain support from close families, our participants indicated that they often obtain emotional support from their relatives and old friends although it may be fading. For example, P6 and P9 received care from daily greetings and conversations with old friends, and P8 shared his daily frustrations with her younger siblings and another older drifter in the same hometown but migrating to another city.}

\subsubsection{Network Support}

\review{Network support refers to the development of a sense of belonging among people with similar interests and concerns, reinforcement, and expansion of a support seeker's social connections \cite{cutrona1992controllability}. It involves spending time with recipients, offering access to new companionship, and reminding the recipients about the availability of companionship. For our participants, 
although new friends and neighbors in the least close convoys seemed to meet the criteria of network support by spending time together, sharing similar hobbies and identities, and broadening new relationships, they were described by our participants as `\textit{temporary relationships based on politeness and equality}', thus providing a very limited sense of belonging. However, local information provided by these relationships might reduce their sense of drifters and improve their belongingness in the host communities (P2,9). Besides, although children in the closest convoy also meet the criteria, they did not provide a sense of belonging for older drifters, because `\textit{children's home would never be my home}' (P3,5). }

\review{Compared to the limited sense of belonging to the new places, all participants expressed their attachment and belongingness to their hometowns, and planned to return to their hometowns in the future when their children did not need them (e.g., for caregiving grandchildren). This strong obsession was supported by their spouses and parents in the closest convoy and old friends and relatives in the closer convoy. For them, returning to hometowns meant familiar relationships (P1-8,11,13), rich social activities (P6,9), `root' traditions and cultural identity (P5,8,9), and freedom (P4), which were their true sense of belonging. }

\subsubsection{Esteem Support}

\review{Esteem support is defined as a `particular type of emotional support that is provided with the intent of enhancing how others feel about themselves and their attributes, abilities, and accomplishments' \cite{holmstrom2012helps}. Our participants expressed that they received insufficient esteem support after migration, mainly coming from their new neighbors and friends in the least close convoy, who had similar experiences with them (e.g., retirement (P13), age-related physical inability (P9), loneliness (P6)). Given the similar experiences, these local friends could understand their situations, and provide them with empathy, support and recognition. Meanwhile, although most older drifters provided great support to their children (e.g. helping babysit), our participants reported that they seldom received appreciation and recognition from them. The main reason was the deeply rooted Chinese Confucianism culture of parents' default responsibilities in supporting the next generations \cite{zhang2023effect}. All participants expressed that they were willing to take these things as their responsibilities, even though their efforts were often taken for granted and received nothing in return.}

\subsection{Roles of ICT in Older Drifters' Convoy}
\label{ICTrole}

ICT has been deeply involved in the social convoys of older drifters. We summarized ICT tools used by our participants in different convoys and their corresponding roles in \autoref{Roles of ICT in Older Drifters' Convoy}. The descriptions of these ICT tools can be found in \autoref{technologies specific to the chinese context} in Appendix. We now elaborate on how these tools work in mediating older drifters' convoys. 

\begin{table*}
    \caption{\ying{Roles of ICT in Older Drifters' Convoy}}
    \label{Roles of ICT in Older Drifters' Convoy}
    \begin{tabular}{p{1.65cm}p{1.95cm}p{3.4cm}p{1.65cm}p{7.35cm}}
        \toprule \textbf{Convoy} & \textbf{Tools} & \textbf{ICT in Connections} & \multicolumn{2}{l}{\textbf{ICT in Support}} \\
        \midrule
            Closest & Tel. & Maintain old connections & tangible & babysitting, full filial piety\\
            - close family & WeChat & - inter-link connections & informational & object of tutoring, assistance and sharing \\
            ~ & Videoconf. & - organize social activities & emotional & remote connect with spouse \\
            ~ & Little Red Book & - share social activities & network & remote connect with parents and spouse \\
            ~ & Headlines, etc. & ~ & esteem & - \\
        \midrule
            Closer & Tel. & Maintain old connections & tangible & communication tools \\
            - relatives & WeChat & - inter-link connections & informational & access hometown information, object of tutoring, etc. \\
            - old friends & TikTok, etc. & - cross-platform connect & emotional & make care \& concern expressive, visible \& propagated \\
            ~ & ~ & - organize social activities & network & remote connect with old friends \& relatives, belongingness \\
            ~ & ~ & - share social activities & esteem & -\\
        \midrule
            Least Close & WeChat & Create new connections & tangible & communication tools\\
            - new friends & TikTok, etc. & - propagate connections & informational & access local information, object of tutoring, assistance, etc.\\
            - neighbors & ~ & Maintain old connections & emotional & - \\ 
            ~ & ~ & - inter-link connections & network & group and maintain by commonality, belongingness\\
            ~ & ~ & - organize/share/record, ... & esteem & communication tools \\
        \bottomrule
    \end{tabular}
\end{table*}

\subsubsection{Promoting Propagated, Inter-linked Social Connections}

\review{Basically, ICT tools, particularly social media such as WeChat and TikTok, helped older drifters navigate transitions in convoys and promote the construction and maintenance of social connections. According to our participants, ICT-mediated social interactions were very common in their daily lives. They often shared and recommended social media accounts (P6) to each other, and invited others to existing social groups (P2), through which social connections flowed and spread among them, improving their social engagement and participation in the host communities.
Meanwhile, group functions of many ICT tools maintained inter-linked connections of older drifters. Across our interviews, our participants presented multiple groups they engaged, such as kinship groups (P1,2,6,8), friend groups (P1,6,8,9,), groups for re-education/re-employment (P3,5,9), groups with common identities (P2,3,8,9,11,13) or hobbies (P4,5,9), etc. The group way promoted inter-linked connections and social participation. 
In addition, ICTs also provided cross-platform communication functions for our participants to connect with more convoy members, offering an approach of reconnecting with fading or lost connections in the remote closer convoy (P6,9).}

\subsubsection{Supporting Organizable, Recordable, and Shareable Social Participation. }

\review{ICT supported the planning, recording, and sharing of social participation, enabling our participants to stay connected and engaged in both new and old convoys. Firstly, ICT tools, especially WeChat groups, served as platforms for them to communicate, plan and organize social activities in host communities. For instance, P8 and P13 told us they often used WeChat groups to make appointments with other friends in advance for weekend outings; Some also used ICT to connect with relatives in their hometown to coordinate specific tasks, such as preparing for Tai Chi competitions (P5) and taking care of their parents collaboratively (P5,7,8). Besides, ICT also served as a recording tool for common activities, preserving and sharing memories of social activities. Our participants told us they often used functions such as group photo albums (P13) to record their interactions and experiences in the new places (P2,5,7). Platforms, such as TikTok, WeChat Moment and WeChat groups, allowed participants to share these recordings with their friends (P4-6,9) and engage in group chatting (P1,2,4) in various convoys. This process also created a sense of `indirect participation' (P6), making them feel connected with remote connections.}

\subsubsection{Navigating Conflicts and Burdens}

\review{ICT provided opportunities for older drifters to navigate conflicts with their intergenerational families, through fostering collaboration, knowledge acquisition, and intergenerational connections among families. 
Our participants expressed that they often organized offline activities and shared recordings of children in WeChat groups with other older drifters. This co-grandparenting approach eased burdens and added joy to the babysitting experience. Meanwhile, ICT enhanced knowledge acquisition, allowing grandparents to learn and update their knowledge of babysitting, which indirectly reduced the possible babysitting-related challenges. 
For instance, P13 often used Little Red Book for learning baby food recipes and Headlines and WeChat public accounts for learning child psychology and education knowledge. She told us these learning processes made her feel babysitting to be very interesting. In addition, leaving hometowns often meant older drifters lacked the opportunity or ability to care for their parents far away. In such cases, ICT could compensate for the lack of in-person family duties to some extent through online communication and sharing (P2,5,7).}

\subsubsection{Improving Information Accessibility} 

\review{ICT promoted older drifters' information accessibility from convoys in both host communities and hometown, which was helpful for them to navigate the challenges they might encounter after migration, increase their engagement in the host community and foster attachment to hometowns.
To be specific, ICT helped our participants connect and engage in the new environment by providing access to information about the host community. Our participants told us they usually gathered daily life information from WeChat groups of local retirees (P2) or learned about local elderly assistance and volunteer organizations from WeChat groups of local cycling enthusiasts (P9). Accessing local information reduced their sense of being a drifter.
Meanwhile, ICT supported participants in receiving hometown information from old connections, which strengthened the connection and attachment to their previous environments and provided a sense of `indirect participation' (P2).}

\subsubsection{Making Concern and Care Visible and Propagated. }

\review{ICT promoted the emotional expression of older drifters in social convoys by making their care and concern expressive, visible and propagated. Specifically, ICT tools, such as WeChat video/voice calls, WeChat groups, TikTok, etc, provided places for older drifters to express their concern and care remotely. The indirect interaction functions, such as, `follow', `Likes', `Commenting' etc., were particularly highlighted by our participants as more convenient and comfortable ways to express their care and attention to others. 
Additionally, ICT also made concern and care in connections propagated through a group/public effect. Interactions often occurred in groups or public spaces, which created a propagation effect between connections, such as the spread of greetings within and between WeChat groups and following the comments under posts. }

\subsection{Expectations of ICT-supported Convoys of Older Drifters}
\label{challenges}

\review{While our participants perceived benefits from ICT in maintaining or creating connections as well as providing support, they also reported a series of challenges and expectations for better ICT-supported social connection and support.}

\subsubsection{Building Trustworthy Relationships} 

Our participants reported difficulties in establishing new connections due to limited mutual acquaintances, reduced opportunities for face-to-face interactions, and a lack of trust within online environments. Presently, they often rely on recommendations from mutual contacts and offline engagements with individuals sharing similar interests, backgrounds, employment statuses, and educational pursuits to forge connections in the new environments. However, in instances where mutual connections were absent, the presence of like-minded individuals or peers of similar age was scarce, face-to-face interactions were minimal, and opportunities for employment or education were unavailable, they encountered significant challenges in establishing ties within the host community. In fact, these situations were very common among our participants.

According to our participants, they had few opportunities to create new connections through offline and online interactions. In particular, the rebuilding process was even harder due to the mistrust stemming from the aging-related factors and tension caused by the increasing fraud cases among older adults. P1, for instance, considered that online connections often would like to obtain some benefits from her, such as selling something, and she thus rejected to be online connected: `\textit{I have mental pressure. I don't know if they are real, or if they could be some kind of scammer.}'

\review{
Given these challenges, our participants expressed their desire for ICT tools that could help them build trustworthy connections in the new environment. Specifically, they hoped to find friends from their hometown (P3,4,7), meet like-minded individuals (P6), and connect with potential friends in the local communities or beyond (P7,9). Our participants expressed that interacting with people from their hometown brought a greater sense of warmth and increased trustworthiness (P1-4,6,8,12), with fewer language and cultural barriers. However, they had yet to discover specific online platforms or channels for this purpose.}







\subsubsection{Supporting Authentic and stress-free Interactions}

\review{Existing ICT tools facilitated the transition of old connections from offline to online, which to some extent promoted the re-connection of our participants with their old relationships and lives. However, there were still challenges reported by our participants in connecting with old relationships, including the lack of authenticity, physical interactions and practical support, which resulted in less intimacy and even fading connections. 
P4, for instance, felt that compared to in-person interactions, videos on platforms like TikTok lacked genuine emotions and often came across as showy, which made him off-putting. Similarly, P6 stated that online interactions had a certain sense of distance and were not as intimate as in-person interactions.}

\begin{quote}
    \textit{`Face-to-face communication is the best way to express one's thoughts and emotions authentically. For example, by inviting a friend to my house today, I can directly perceive whether s/he is happy or not. However, communications through  WeChat and TikTok couldn't support these, leading to a sense of distance. On these platforms, we tend to respond politely, e.g. following or liking. This lacks authenticity. I prefer genuine and authentic communication.'} (P6)
\end{quote}

Meanwhile, our participants also pointed out the lack of physical experiences in online interactions, which made it hard to maintain intimate relationships. P1 shared that she used to spend time with friends in her hometown, but recently their connections had grown distant due to online communication. Similarly, P9 believed that although online interactions were convenient, they couldn't replace real-life interactions with friends in his hometown, such as having meals together or going out for drinks. In addition, our participants also expressed that online interactions made it difficult to express difficulties or provide practical support when their friends and relatives needed it. They therefore hoped that ICT tools could help make interactions more authentic and warm.

In addition, for older drifters, sharing their personal lives was an important way to stay connected with friends and families who were far away. However, our participants expressed pressure on existing social sharing platforms, whether they were public platforms such as TikTok or semi-public platforms such as WeChat Moments. For instance, P6 expressed that when sharing her happy life online, she often worried that others might perceive it as showing off. This kind of mental stress reduced their willingness to share, affecting their abilities to reconnect with relationships.

\subsubsection{Promoting Harmonious Family Relationships}

\review{Our participants reported a series of digital grandparenting conflicts, such as the contradiction between using smartphones and setting the right digital role models for young children (P7,13), difficulties in interacting with their adult children (P2,4,6-9,11), etc., and hoped there were technologies which could negative these intergenerational conflicts, and promote technology-based interactions within families. 
For example, in the process of babysitting, especially with young children, frequent smartphone use may set a poor example. P13 mentioned that she only used her phone at night to avoid her grandchild imitating this behavior. P7 expressed a similar viewpoint, explaining that whenever he played with his phone, his grandchild came to snatch it, so he could only use it after the child fell asleep. They therefore desired a technology that they could use together with their grandchildren.}

At the same time, our participants also mentioned difficulties in getting along with their adult children, such as generational gap (P4,8) and lack of communication about their practical difficulties or emotional needs (P3,5-9). P2 and P8, for instance, mentioned that their children came and went hurriedly every day, with limited conversation with them. Even when they had a chance to chat, they usually found themselves detached from the conversation, not actively participating in the discussions between young people. Furthermore, in traditional Chinese culture, parents usually played the role of self-sacrifice. When facing problems such as loneliness (P3), declining memory and health issues (P5), worrying about being disliked by their children (P6), and other troubling aspects of life (P5,7-9), they often chose to conceal them and reported only what was good to avoid causing worries from their children. As a result, they hoped for effective ways to enhance communication with their families, whether or not technology was used.

\subsubsection{Improving Information Management}

\review{When ICT played a crucial role in providing information support and fostering relationship connections in the lives of migration, our participants presented challenges in information management and selection. Most of our participants told us that although they relied on local information and hometown updates to navigate their challenges during migration and develop belongingness, they often felt exhausted when dealing with a large amount of irrelevant information from their connections. Therefore, they expected that ICT could assist in information selection, whether sending information to others or receiving information from others.
P9, for instance, described that when learning from WeChat groups, he often felt overwhelmed by the sheer volume of messages, and hoped there was a technological solution to filter them, such as interest-based grouping and recommendation.
}

\begin{quote}
    \textit{`Everyone has different areas of expertise or interests. When they share things they are interested in, you can decide to look at them or not. However, currently all information is piled together, which often makes me confused. It would be great if it is organized based on my interests.'} (P9)
\end{quote}

\section{DISCUSSION}

\finalmark{
This paper aims to deeply examine the social connection and support of late-life migrants. Taking older drifters in China as study objects, we unpacked older drifters' social connections and support through the lens of the social convoy.
In this section, we elaborate on how this social convoy-enabled understanding contributes to existing research on technology-mediated social support of late-life migrants as well as the broader migrant community. Following that, we propose design considerations for future technologies.  }

\subsection{\finalmark{Unpacking Social Support of Late-life Migrants: A Convoy Perspective}}

\finalmark{Through the interview study with older drifters in China, our study presented the ecological social convoy of late-life migrants. In particular, drawing on Kahn and Antonucci's social convoy model \cite{rl1980conboys}, we categorized the social connections in and around late-life migrants into three tiers based on their closeness level with late-life migrants, and elaborated on the dynamic structure of these connections before and after migrations (see Section \ref{structure} for more details). We also illustrated the corresponding roles within each convoy and the support they offered (see Section \ref{role} for more details).}

\finalmark{
Further, we emphasized the dynamic, personal and situational features of older drifters' social convoy. Specifically, the social convoys of older drifters are dynamic in terms of composition, size and quality during migration. The closest convoys, for instance, were relatively stable, although some cases impacted the quality of relationships, e.g., conflicts between babysitting and fulling filial piety, separating from their spouses, etc. Instead, the closer and least close convoys underwent great changes in composition, quantity and quality during migration. In hometowns, late-life migrants lead relatively stable social convoys because of long-term living. Yet, the migrating process breaks up these connections to some extent. Late-life migrants therefore need to re-develop connections in the new places for obtaining the necessary social support. }

\finalmark{Meanwhile, the social convoy of older drifters was influenced by specific personal (e.g., age, retirement, socio-economic status, and personality) and situational factors (e.g., economic development, societal expectations, and traditional culture). For example, aging often caused more difficulties for older drifters in adapting to new language and culture contexts \cite{da2015later} and technologies \cite{friemel2016digital}, and strong dependence on old connections \cite{hiller2004new};
Socio-economic status influenced their social acceptance and sense of belonging in the new environment;
Personalities such as personal interests often play an important role in building new relationships, serving as a bond to construct new connections.
The situationality mainly reflected the influence of social awareness and culture on the migrating behaviors and convey structure. In our case, traditional Chinese family culture and social stereotypes (e.g., parents had the responsibility of helping their children take care of the next generation \cite{fan2006confucian, zhang2006will, lam2006contradictions}) were one main reason for the presented structure of social convey.}

\finalmark{
Additionally, we demonstrated the role of ICT in late-life migrants' social convoy, including maintaining old connections, facilitating the re-rebuilding of new connections, promoting social engagement and participation in new contexts, regulating negative emotions, etc. (see details in Section \ref{ICTrole}). 
This social convoy-enabled understanding provides technology designers and policymakers with a more systematic understanding of the social support status of late-life migrants. Based on this understanding, more targeted support strategies could be designed.}

\subsection{\finalmark{Enabling the Social Convoy of Migrant Community}}

\finalmark{As the number of migrant communities continues to grow \cite{refugee, mcauliffe2020report}, research on migrants has drawn increasing attention in HCI and CSCW fields. Within existing literature, understanding the social connection and support of migrants in new environments and providing appropriate ICT-mediated support strategies have been the focal points of scholars (e.g., \cite{xu2017community, hussain2020infrastructuring, rao2016asian, hirsch2004speakeasy, wyche2012we}). Our paper contributes to this effort by offering empirical insights into the social support networks of late-life migrants, which are largely underexplored in existing literature. Specifically, we introduce the social convoy model \cite{rl1980conboys} into this field, unpacking the social connections and support systems of late-life migrants through this model. Compared to previous studies that analyzed migrations' social support from a relatively static perspective \cite{almohamed2019rebuilding, hsiao2018technology}, the social convoy perspective offers a more dynamic and systematic understanding of individuals' social relationships and transitions, as well as the factors influencing and influenced by social convoy. }

\finalmark{We, therefore, propose incorporating the concept of the social convoy into migration studies in HCI with the following envisioned scenarios. Basically, employing the structured social convoy allows us to systematically examine the dynamics and mobility of social support of migrants, including transitions from one stable state to another, disruptions within convoys, methods of reconstruction (how and by whom), etc. Such insights could provide valuable empirical evidence for designing future ICT-mediated social support systems, shedding light on more targeted strategies for fostering resilient social connections and support networks of migrations.}

\finalmark{Secondly, while the data-driven approach (generally referring to the approach of making decisions and conducting activities based on the analysis of data rather than solely relying on intuition or personal experience \cite{wolf2010data}) has become the mainstream trend in contemporary society, scholars have noted the socio-technological gap \cite{sun2023data} faced by data-driven support systems. One significant reason for this lies in the tension between structured and institutionalized data-driven logic and the complicated social support landscapes \cite{sun2023care, sun2023data, chen2023maintainers}. The social convoy concept emerges as a potential framework for quantifying the complicated support networks within migration contexts, thereby supporting the work of data-driven social service systems. }

\finalmark{And thirdly, practitioners and scholars in public health and social services, have pointed out that visualizing the social support of care receivers could serve as a form of social and psychological support for care and support receivers, potentially enhancing their well-being \cite{goode1995impact, caremap}. In many healthcare settings, quantifying care and support networks (e.g., \cite{guo2022caremap, mclachlan2016using, dunn2006development}) has been utilized as an effective caring approach. In the scenario of providing social support for migrations, we believe that the design of visualization tools based on the social convoy concept also holds the potential to become an effective means of social support.}

\subsection{Design Considerations}

\finalmark{Our study suggested that despite the widespread use of various ICT tools and platforms among our participants in their daily lives (as shown in \autoref{technologies specific to the chinese context}), most of these technologies utilized were generic applications. While these tools have played a crucial role in facilitating social connections for our participants, they do not fully accommodate the unique characteristics and needs of this demographic. Our participants reported various challenges in utilizing these technologies to establish social connections, such as difficulties in navigating overwhelming information and the inability to forge meaningful connections through existing social media platforms, along with their expectations for future technological advancements (see details in Section \ref{challenges}). In existing HCI and CSCW literature, there has been growing attention towards technological innovations aimed at supporting migrant populations (e.g. \cite{wyche2012we, pan2013exploration, gan2020connecting}). However, attention towards late-life migrants remains limited. Considering the increasing number of late-life migrants \cite{Xinhua2018report, UNFPA2019report}, we encourage large investments in technology interventions tailored specifically to this demographic. Our study presented a deep understanding of the social connections and support of late-life migrants in China, which we believe is valuable in informing future technologies for better supporting late-life migrants. We now draw on our findings to discuss the design implications. }

\finalmark{\textit{Supporting Elderly-Friendly Connections within Local Communities. }As illustrated in Section \ref{role}, assisting late-life migrants in establishing social connections in the new environment and enhancing their inclusivity and engagement within local communities is crucial for tangible, informational and network support, directly impacting their life quality. However, late-life migrants in our study encountered a series of practical challenges in building connections within local communities, potentially stemming from psychological, cultural, and language barriers (see details in Section \ref{challenges2}). 
In HCI and CSCW, increasing attention has been paid to technologies that mediate local community building, such as \cite{fedosov2023zuri, fedosov2022supporting, fedosov2021designing}. We believe these technological directions could provide opportunities to bridge connections between late-life migrants and local communities. Additionally, our participants frequently report various challenges in their daily lives in local communities, such as lack of access to local information or inconvenience in using local social service systems. Some of them arise due to the absence of suitable mediums, while others result from the complexities of infrastructuring work \cite{chen2023maintainers}. We therefore advocate for significant technological efforts to facilitate a more convenient local life for late-life migrants, emphasizing the need for elderly-friendly initiatives.}

\finalmark{\textit{Developing Trustworthy and Quality Online Space for Late-life Migrations. }
Besides connections with local communities, ICT-mediated online participation also serves as a crucial component of social engagement for late-life migrants. Our study revealed that, despite the considerable potential of ICT in facilitating the establishment of online social engagement, existing tools often fall short of creating quality social links. One primary reason is the lack of trust in online connections among late-life migrants. In the case of Chinese older drifters, apart from the age-related technological distrust \cite{sun2014being}, the broader social context, such as the prevalence of internet fraud targeting older adults \cite{ekoh2023understanding}, further exacerbates this mistrust. Addressing how to design and develop trustworthy channels for online connections is thus a critical challenge for ICT-mediated social support systems for late-life migrants. 
The trust issue in both online and local communities has been largely explored in HCI and CSCW \cite{kang2016role, fan2018online}. While various technical solutions have been proposed, these may not always be effective for late-life migrants due to the potential digital divide \cite{van2006digital}. Echoing with prior literature \cite{sun2014being}, our suggestion is to seek methods for establishing trust within the social and cultural environment of late-life migrants. For example, in our studied Chinese environment, platforms initiated by government agencies, rather than technological solutions, are the most trusted ones among the elderly population \cite{chen2022exploring}. Additionally, based on feedback from our participants, connecting with individuals who share common characteristics (such as identities, cultural backgrounds, or interests) can help increase the acceptance and usage of online social interactions among late-life migrants.}

\finalmark{\textit{Value-sensitive Technology Design. }
In aging societies such as China, resources for employment and education tailored to the elderly population are limited \cite{heikkinen2013transnational}. The process of migration disrupts the existing social connections of these elderly individuals, resulting in even fewer re-employment and education opportunities for them. As a result, late-life migrants often hold limited opportunities to realize their personal and social value in their new environment. This, in turn, causes a lower quality of life and well-being. 
This is a socio-economic issue that necessitates structural adjustments within the social service system. From a technological design perspective, in addition to fostering connections between late-life migrants and both local and online communities as mentioned above, we further suggest value-sensitive design \cite{friedman2013value} considerations for this population, which we believe holds significant potential in addressing these challenges. For instance, ICT-mediated mutual aid or knowledge-sharing platforms could provide late-life migrants with opportunities to help each other and share their personal experiences, skills or knowledge. Different from public social media platforms (such as TikTok), these platforms specifically targeting late-life migrants are more likely to support resonant and empathic interactions among them, which is expected by our participants. Further, this kind of value-sensitive platform could have highly positive effects on enhancing personal values, and meaningful self-presentation.}

\subsection{Limitations and Future Work}

\review{The primary limitation of our study lies in the diversity of participants. Specifically, although our participants came from various provinces of China, they predominantly resided in Beijing and Shanghai, two biggest cities in China. These demographic characteristics of the participants pose limitations in sample diversity. Meanwhile, most of our participants became late-life migrants due to the need to care for their children or grandchildren. However, in reality, there are diverse reasons for becoming late-life migrants, which our study did not cover. 
In addition, our study was conducted in China, and our participants exhibited strong cultural attributes, which also impacted the generalizability of our research findings. In the future, we will expand our study to more diverse late-life migrant populations, including those with diverse current residences, varied motivations of migration, and diverse social and cultural backgrounds, and provide a more comprehensive social convoy landscape of late-life migrants.}

\section{CONCLUSION}

\review{This paper presents a qualitative study that investigates social connections and support of late-life migrants. Drawing on the social convoy model, we contribute an in-depth empirical understanding of the structure and transitions of late-life migrants' social convoys, the support roles of each convoy, the functions ICT plays, as well as the encountered challenges and expectations of our participants regarding ICT-supported social convoys. Further, we unpack the role of the social convoy in supporting more targeted social support for late-life migrants and broader migrant communities, ending by offering late-life migrant-oriented design considerations.}

\begin{acks}
    We thank all the interviewees for contributing their time and insights. We also thank the anonymous reviewers for helping us to improve this work.
\end{acks}

\balance

\bibliographystyle{ACM-Reference-Format}
\bibliography{main}



\appendix

\onecolumn

\section{Used Technologies}

\begin{table*}[htbp]
    \caption{\ying{Descriptions of Technologies Used by Our Participants}}
    \label{technologies specific to the chinese context}
    \begin{tabular}{p{3cm}|p{14cm}}
        \toprule \textbf{Technologies} & \textbf{Description} \\
        \midrule
            WeChat & a multi-purpose instant messaging, social media and mobile payment app, with a lot of subfunctions such as friends adding, group chat, audio/video chat, moment (life sharing in friends circle),  public accounts (information and article sharing by governments, organizations or individuals), mini-programs, etc.\\
            \midrule
            TikTok & a short video sharing mobile application.\\
            \midrule
            Tencent Meeting & an audio and video conferencing software under Tencent Cloud.\\
            \midrule
            Little Red Book & a social media and e-commerce platform; is sometimes referred to as `Chinese Instagram'.\\
            \midrule
            Headlines & a Chinese news and information content creation, aggregation and distribution platform.\\
            \midrule
            Tencent Video & an online video media platform launched by Tencent.\\
        \bottomrule
    \end{tabular}
\end{table*}


\end{CJK*}
\end{document}